\documentclass[twocolumn,aps,prb,floatfix,showpacs]{revtex4}
\usepackage{graphicx}
\usepackage{dcolumn}
\usepackage{bm}
\usepackage{amsfonts}
\usepackage{amssymb}

\begin{document}
\pagenumbering{arabic}
  \title{Spintronic devices from bilayer graphene in contact to ferromagnetic insulators}
  \author{Paolo~Michetti and Patrik~Recher}
  \affiliation{Institute of Theoretical Physics and Astrophysics, University of W\"urzburg, D-97074 W\"urzburg, Germany}
  \email{michetti@physik.uni-wuerzburg.de, precher@physik.uni-wuerzburg.de}
  \pacs{72.25.-b, 72.80.Vp, 85.75.-d, 85.75.Mm, 85.70.-w}
  
  \date{\today}
  
  \begin{abstract}
    Graphene-based materials show promise for spintronic applications
    due to their potentially large spin coherence length.  
    On the other hand, because of their small intrinsic spin-orbit interaction, 
    an external magnetic source is desirable in order to perform spin manipulation.
    Because of the flat nature of graphene, the proximity interaction with a ferromagnetic insulator (FI) surface seems 
    a natural way to introduce magnetic properties into graphene.
    Exploiting the peculiar electronic properties of bilayer graphene 
    coupled with FIs, we show that it is possible to devise very efficient gate-tunable spin-rotators and    spin-filters in a parameter regime of experimental feasibility.    %
    We also analyze the composition of the two spintronic building blocks in a spin-field-effect 
    transistor. 
  \end{abstract}
  
  \maketitle

  Graphene with its high mobility~\cite{geim2007} and potentially long spin lifetimes, 
  is an attractive material for spintronics. 
  In particular, spin relaxation lengths on the order of micrometers have been observed~\cite{tombros2007}, 
  together with spin relaxation times of hundreds of picoseconds, which are still believed to be limited by extrinsic 
  impurities~\cite{tombros2008, ertler2009}.  
  More recent experiments reported the measurement of a spin lifetime up to $1$~ns in graphene 
  and even of several nanoseconds in bilayer graphene (BG)~\cite{han2011, yang2011}.
  %
  Moreover, tunnel-injection of spin into graphene has been recently achieved using Co ferromagnets, 
  with the observation of the largest non-local magnetoresistance of any material~\cite{han2010}. 
  Graphene quantum dots have been also identified as an ideal host for spin qubits~\cite{trauzettel2007,recher2010}.

  The reason for such favorable spin properties is the small spin-orbit coupling (SOC) and the weak hyperfine 
  interaction with the underlying nuclear spin system~\cite{yazyev2008, fischer2009}. 
  The SOC in single layer graphene has been predicted to be on the order of $10^{-3}$-$10^{-2}$~meV~\cite{min2006,huertas2006,gmitra2009}.
  %
  On the other hand, this weak SOC constitutes a severe limitation for spin manipulation in 
  conventional spintronics devices like the Datta-Das spin-field-effect transistor (SFET)~\cite{datta1990}.

  An alternative strategy is offered by contacting graphene with 
  a ferromagnetic insulator (FI), giving rise to an exchange proximity 
  interaction (EPI)~\cite{semenov2007, haugen2008, semenov2008}.
  EPI results from the Coulomb exchange interaction between $\pi$ states in graphene 
  and the magnetic ions on the FI surface.
  Ideally, the EPI---being short-ranged---affects only a graphene layer in direct contact with the 
  FI and acts like an effective Zeeman field, superimposed on the original BG Hamiltonian~\cite{semenov2008}.

  %
  %
  %

\begin{figure}[bp]
    \centering	
    \includegraphics[width=8.5cm]{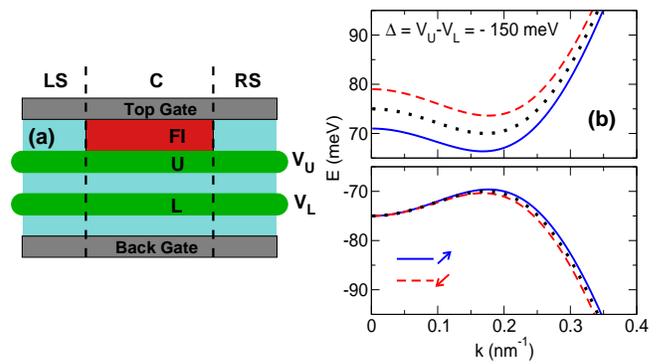}
    \caption{(Color online)
      (a) Setup of the double gate BG with FI used as a spacer between the U layer and the top gate.
      (b) Dispersion curve of biased BG, showing the typical Mexican hat behavior, subject to EPI.
      A gate-bias $\Delta=-150$~meV is imposed between the BG layers. 
      The full blue lines are $\nearrow$-spin polarized bands (concord to the FI magnetization axis), 
      while the dashed red lines are $\swarrow$-spin polarized.
      The dotted lines represent the normal non-magnetic BG dispersion, as used for the LS and RS leads.
    }
    \label{fig:devdisp}
  \end{figure}

  Here, we theoretically study transport through BG in a double gate configuration, on which a FI is used as
  spacer between the upper (U) layer and the top gate, giving rise to EPI, as shown in Fig.~\ref{fig:devdisp}(a). 
  The gate bias $\Delta$ is used to impose a semiconducting gap and to localize the low-energy region
  of the conduction band on the U or the lower (L) layer (depending on the {\it sign} of $\Delta$). 
  Indeed, a tunable semiconducting gap up to $250$~meV with the application of a gate bias has been 
  demonstrated~\cite{castro2007, Oostinga2008, zhang2009}.
  Consequently, it is possible to electrically control the effective Zeeman field for electrons in the 
  conduction band, turning ON or OFF the device.
  We show, in particular, that the device can act either as a spin-rotator (SR) or as a spin-filter (SF). Finally, we propose and analyze the combination of these two spintronic building-blocks within a SFET.

  \section{Setup and scattering problem\label{sec1}}
  The Dirac Hamiltonian describing BG near the K point is~\cite{mccann2006}
  \begin{equation}
    H_0 = - \frac{\Delta}{2} \tau_z + v_f (\sigma_x p_x + \sigma_y \tau_z p_y) + \frac{t_\perp}{2}(\sigma_z+\sigma_0)\tau_x,
  \end{equation}
  with $\vec{\sigma}$ and $\vec{\tau}$ the Pauli matrices for the  sublattice (A,B) and layer (U,L) degrees of freedom, 
  $\Delta$ the potential energy difference between the U and L plane,  $t_\perp=0.39$~eV the inter-layer hopping 
  parameter~\cite{nilsson2008}, and $v_f\approx 10^{6}$~m/s~\cite{castroneto2009}.
  The Hamiltonian acts on the the spinor   
  $\Psi = \left(\chi_{B'}, \chi_{A'}, \chi_{A}, \chi_{B}\right)$, with $A$ and $B$ on the U layer, 
  $A'$ and $B'$ on the L one.

  When the U layer is placed in direct contact with the FI surface, 
  it introduces a Zeeman field affecting the U part of the Dirac Hamiltonian~\cite{semenov2007, haugen2008, semenov2008}
  \begin{equation}
    h_m= -\frac{E_z}{2} \hat{m} \vec{s} \left(\tau_0-\tau_z\right), 
    \label{eq:EPI}
  \end{equation} 
  where $\hat{m}=(m_x,m_y,m_z)$ is FI's magnetization axis, 
  $\vec{s}=(s_x,s_y,s_z)$ are the spin Pauli matrices and
  $E_z$ is the effective strength of the EPI (absolute magnitude of the Zeeman splitting).

  We consider now a central (C) barrier region of length $L_C$, 
  made of BG subject to EPI and described by $H=H_0+h_m+ U_0$, with $U_0$ a possible potential shift, while the left side (LS) and the 
  right side (RS) leads are semi-infinite normal BG described by $H_0$.
  In LS and RS regions, the dispersion curves are degenerate in the spin degree of freedom.
  In the C region, a spin-splitting $E_z$ arises between the spin components which are parallel ($\nearrow$) and 
  antiparallel ($\swarrow$) to $\hat{m}$, see Fig.~\ref{fig:devdisp}.
  A detailed description of the eigenstates of $H_0$ and $H$ is given in Appendix \ref{app1}.

  In the present paper we use $E_z=8$~meV, close to the estimation of Ref.~\onlinecite{haugen2008}, 
  and, when not stated otherwise, $|\Delta|=150$~meV and a temperature of $T=1$~K.
  In Fig.~\ref{fig:devdisp}(b), we show the lowest conduction and valence bands in 
  the C region (full, dashed lines) for $U_0=0$ and in the leads regions (dotted lines).
  The spin-splitting for conduction and valence bands is proportional to the localization of their 
  respective states on the U layer (see Fig. \ref{fig:minimo}). 
  Therefore, inverting $\Delta$ will invert the spin splitting 
  for electrons and holes.  
  \begin{figure}[tbp]
    \centering	
    \includegraphics[width=8.cm]{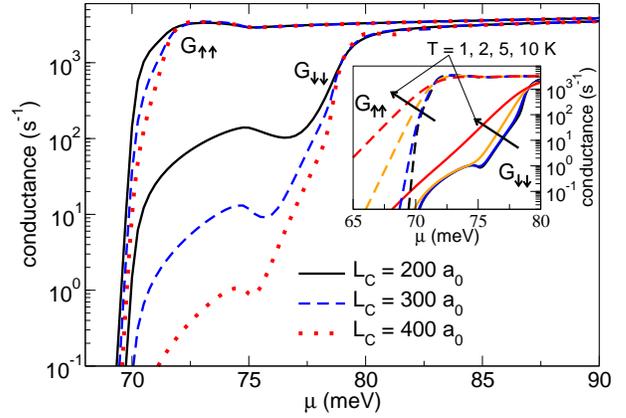}
    \caption{(Color online) Conductance of spin up and spin down carriers as a function of chemical 
      potential $\mu$ for a BG device acting as a SF calculated for increasing values of $L_C$.
      A FI is in contact to the U layer of the BG, giving rise to an EPI with $E_z=8$~meV.
      The gate-bias is $\Delta=-150$~meV and a potential shift of $U_0=5$~meV is applied in the C region.
      In the inset, we show the spin-resolved conductance of the SF with $L_C=400a_{0}$ calculated 
      for different temperatures.
    }
    \label{fig:SF}
  \end{figure}

  The system is assumed to be invariant under translations along $Y$ and the scattering is elastic. 
  Therefore $k_y$ and $E$ are conserved quantities. 
  We briefly outline here the procedure used to calculate the transmission through a single barrier by spinor matching 
  and the conductance of the system, which has been introduced in Ref.~\onlinecite{michetti2010b}.
  For a given $E$ and $k_y$, solving the Hamiltonian $H_0$($H$), we obtain an analytical 
  description of the spinors of the propagating and evanescent modes in the LS and RS (C region), 
  which contribute to the scattering state.   
  In the RS and LS regions, disregarding the spin, there are four possible values 
  of the wavevector $k_x$ compatible with a given $k_y$ and $E$: $k_x$ and $-k_x$, which are propagating modes, 
  $\tilde{k}_x$ and $-\tilde{k}_x$, which correspond either to propagating or to 
  evanescent modes having a finite imaginary part~\cite{barbier2009}.
  In the central part---due to the EPI---the secular equation for $H$ leads to spin-dependent 
  solutions of the wavevector $k_x=\alpha_n$ with $n=1$, $2$ \dots $8$, described by 
  the spinors $\Psi_{\alpha_n}(x)$, which are eigenstates of $H$.

  For an incoming particle of wavevector $k_x$, $k_y$ and spin-polarization $\vec{s}$ (a vector describing the up 
  and down spin components with respect to the $Z$ axis), 
  we solve the linear system determined by imposing the continuity of the scattering state at $x=0$ and $x=L_C$.
  This fixes the output transmission(reflection) coefficients $t_{\uparrow}, t_{\downarrow}, 
  \tilde{t}_{\uparrow}, \tilde{t}_{\downarrow}$($r_{\uparrow}, r_{\downarrow}, \tilde{r}_{\uparrow}, 
  \tilde{r}_{\downarrow}$) 
  for the allowed $k_x$ and $\tilde{k}_x$ modes in up or down spin orientation. 
  We define the spin-resolved transmission probability $T_{\lambda,\lambda'}$ as the sum of transmission probabilities in 
  all the outgoing propagating modes ($|t_{\lambda}|^2$, $|\tilde{t}_{\lambda}|^2$), 
  calculated for $\vec{s}$ compatible with a spin-polarization of the incoming particle $\lambda'$.

  We calculate the conductance of the ballistic system in linear response.
  The 2D  two-terminal conductance is 
  \begin{equation}
    G_{\lambda,\lambda'} = \frac{g e^2}{(2\pi)^2} \int\int dk_x dk_y T_{\lambda,\lambda'}  v_x \frac{df(E-\mu)}{dE},
  \end{equation}
  with $g=2$ (accounting for the valley degeneracy), $f(E)$ is the Fermi-Dirac distribution, 
  $v_x$ the group velocity along the transport direction, $E$ the particle energy, 
  and $\mu$ the electrochemical potential.

  \begin{figure}[tbp]
    \centering	
    \includegraphics[width=8cm]{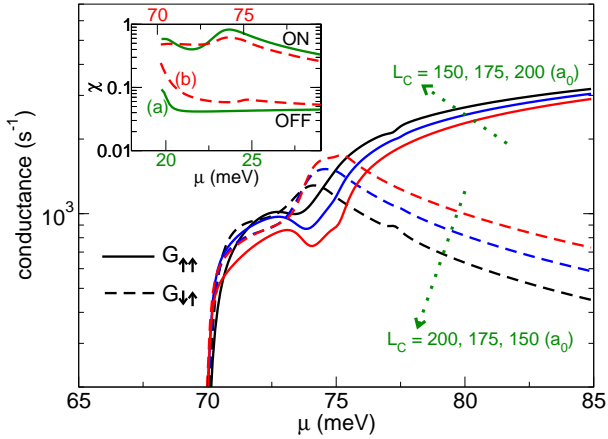}
    \caption{(Color online) Conductance with transmission into spin up $G_{\uparrow\uparrow}$ and 
      spin down $G_{\downarrow\uparrow}$ carriers for a BG device acting as a SR in ON state, 
      feeded by $\uparrow$-polarized electrons.
      In the inset the spin-flip conductance fraction 
      $\chi=G_{\downarrow\uparrow}/(G_{\downarrow\uparrow}+G_{\uparrow\uparrow})$ is 
      shown for ON ($\Delta<0$) and OFF ($\Delta>0$) state of the SR for $L_C=175 a_0$.
      The dashed curves, data set (a), are calculated with $|\Delta|=40$~meV for $L_C=200 a_0$ . 
    }
    \label{fig:SR}
  \end{figure}

  \section{Spin-Filter}
  We now analyze the behavior of the device as a SF.
  In particular, the device acts on unpolarized incoming particles, filtering
  the component antiparallel ($\swarrow$) to $\hat{m}$.
  In Fig.~\ref{fig:SF}, we show the spin-resolved conductance of the device as a function of $\mu$, 
  where we choose $\hat{m}$ along $Z$ and a potential shift 
  of $U_0=5$~meV. 
  When $\mu$ falls between the spin-splitted bands (between $70$ and $78$~meV) in the C region, 
  $G_\downarrow$ is exponentially suppressed as a function of $L_C$ with an average 
  effective decay length of the order of $50$~nm, while $G_\uparrow$ does not vary.
  This behavior is due to the fact that, in this energy range, 
  transmission of spin down particles occurs through evanescent modes, which 
  exponentially decay (in $C$).
  Thus, in the spin-splitted C region and for $T\ll E_z/k_B $, the device acts as an efficient SF($\uparrow$), 
  i.e. it lets pass only current with $\uparrow$ spin polarization.
  Such a SF can be used to generate a spin polarized current out of an unpolarized one.
  Or reversely, it can be used as a spin analyzer which detects the degree of spin-polarization of charge 
  carriers. 
  This possibility will be exploited later in this paper. 
  In the inset of Fig.~\ref{fig:SF}, we show the spin-resolved conductance of the BG SF for increasing 
  temperature in the range of $1$ to $10$~K.
  As expected, thermal excitations degrade the SF efficiency.
  In particular, the value of $E_z$ imposes a maximum operating temperature for the device of $T\approx E_z/8 k_B$, 
  which corresponds to about $12$~K, for $E_z=8$~meV.


  \section{Spin-rotator}
  We have shown in a previous work~\cite{michetti2010b}, that a BG in contact with a FI
  can act as an electric-field switchable SR. 
  The control of spin-rotation with the gate bias essentially depends on the degree 
  of wave function localization on one of the BG layers near the Mexican hat energy dispersion region.
  A useful parameter to characterize spin-rotation is the ratio $\chi$ of 
  the conductance associated with a spin-flipped transmission to the total conductance~\cite{michetti2010b}.
  As shown in the inset of Fig.~\ref{fig:SR},  we can put the device 
  OFF ($\chi_{\rm OFF}\approx6\%$) or turn it ON ($\chi_{\rm ON}\approx60\%)$ by reversing the gate bias.
  The performance of the spin-rotator is basically limited by the finite $\chi$ fraction in the OFF state, 
  due to the non perfect layer localization of electrons contributing to transport.
  With a smaller gate-bias of $|\Delta|=40$~meV~\footnote{Our model does not account for the trigonal warping correction
    that introduces a strong angle-dependence of the transmission for energies as small as $\mu\approx|\Delta|=40$~meV.}, 
  the layer confinement is more effective (see Fig.~\ref{fig:minimo}) and our model predicts  
  a moderate performance enhancement ($\chi_{\rm OFF}\approx4\%$ and $\chi_{\rm ON}\approx80\%$).
  Fig.~\ref{fig:SR} shows the spin resolved conductance in the ON state for a SR feeded by a
  $\uparrow$-spin-polarized source lead, for increasing lengths of the magnetic barrier.
  In particular, for $L=175 a_{0}$ ($a_{0}=1,42 \mathring{A}$ the in-plane nearest neighbor distance), 
  a strong spin-flip resonance is present at $\mu\approx74$~meV, which is 
  well inside the SF operational regime presented in Fig.~\ref{fig:SF}.
  \begin{figure}[tbp]
    \centering
    \includegraphics[width=8cm]{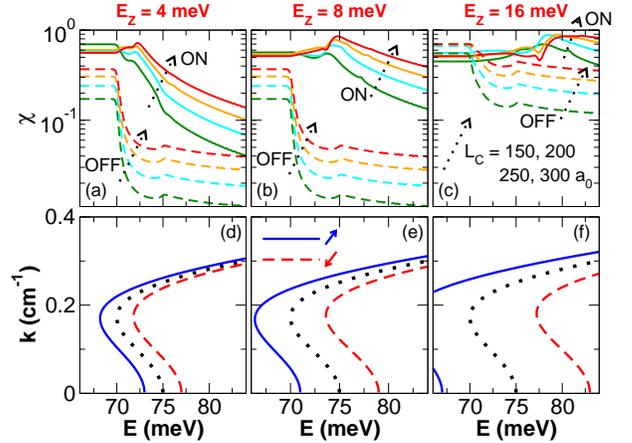}
    \caption{(Color online) In (a), (b) and (c), we present the $\chi$ factor calculated for a SR with different values of the $E_z$ parameter for $L_C$ varying from $150$ to $300~a_0$. 
      For comparison we show in (d), (e) and (f) the corresponding first conduction band dispersion: in a dotted curve that of the leads, 
      while in full and dashed lines the $\nearrow$- and $\swarrow$-spin-splitted bands in the C region of the SR, corresponding to the ON state.  
    }
    \label{fig:Ez}
  \end{figure}
 
  The origin of the spin rotation is easily explained.
  A $\uparrow$-spin-polarized electron is described inside the spin-rotator as the superposition of two components with spin polarization
  $\nearrow$ and $\swarrow$ (polarizations which are parallel and antiparallel to the FI magnetic axes and are eigenstates of the EPI system).
  These two components, being coupled differently to the EPI, travel with a different $k_x$ wavevector and accumulate a 
  phase difference $\Delta k_x \cdot L_C$ in a single crossing of the C region.  
  The phase difference translates into a net rotation of the initially $\uparrow$-spin-polarized electron.
  Due to the complex 2D BG dispersion curve, it is difficult to establish an immediate relation for the spin-flip resonance condition between the parameters 
  $\Delta$, $E_z$ and $L_C$.
  However, for a wide range of $E_z$ values, a spin-flip resonance is observed in the Mexican hat region of the $\swarrow$-spin-polarized band, 
  for $|\Delta|=150$~meV and $L_c \approx 200 a_0$, as shown in Fig.~\ref{fig:Ez}.
  The order of magnitude of $L_C$ is related to $\pi/k_{min}$, where $k_{min}$ is the wavevector corresponding to the minimum of the first conduction band (Eq.~\ref{eq:kmin}).  
  In Fig.~\ref{fig:Ez}, we compare the $\chi_{ON}$ and $\chi_{OFF}$ of three systems with $E_Z=4$ (a), $8$ (b) and $16$~meV (c), 
  for a C region of length $L_C= 150$, $200$, $250$ and $300$ $a_0$.
  We also plot in Fig.~\ref{fig:Ez}(d-f) the corresponding $k_y=0$ dispersion curves for the normal BG (dotted) 
  and for the BG subject to EPI in the ON state, in a full curve for $\nearrow$- and in a dashed curve for $\swarrow$-spin-polarization.  
  In correspondence to the edge of the $\swarrow$-spin-polarized band, transport is often associated with a maximum of $\chi_{ON}$, 
  corresponding to the fact that the majority of the electrons which tunnel through the C region are spin-flipped (i.e. they satisfy $\Delta k_x \cdot L_C\approx \pi$).

  \section{Spin FET}
  We now calculate the total conductance  $G=\sum_{\lambda,\lambda'}G_{\lambda\lambda'}$ 
  for a hybrid setup made by the series of a SF($\uparrow$), 
  a SR(ON/OFF) and a SF($\downarrow$), each one built from BG in contact with a FI with $\hat{m}$ 
  along $Z$, $Y$ and $-Z$, respectively.
  Ideally, the SF($\uparrow$) selects the $\uparrow$-component of the incoming unpolarized electrons, resulting in 
  a spin-polarized current.
  The SR introduces a spin precession which we can turn ON or OFF
  with the gate bias (see inset of Fig.~\ref{fig:SR} and related discussion).
  Finally, the SF($\downarrow$) measures the degree of spin-rotation, because it (ideally) lets pass
  only carriers which have been spin-flipped by the SR.
  Therefore this structure realizes a complete spintronic scheme of creation, manipulation and measurement
  of spin-polarized currents, which does not require spin-polarized leads.   
  The calculation of the transmission is performed by applying the transfer-matrix (TM) formalism, which we briefly review in Appendix~\ref{app2}.
  For each barrier we obtain the corresponding TM, which requires $\vec{r}$ and $\vec{t}$ 
  for the individual scattering problems of a particle, approaching the barrier from the 
  LS or from the RS.
  The TM links the LS modes to the RS modes and therefore is multiplicative, 
  in the sense that the TM of a series of barriers is the ordered product of the corresponding individual TMs. 
  We thereby neglect the contributions of the evanescent modes connecting different scattering regions of the system.
  Therefore we consider the three spintronics blocks separated by a distance of $L=1000 a_{0}$, where 
  any effects of such evanescent modes are negligible \footnote{However, we do include the evanescent modes 
    for the scattering problem of the individual blocks where they are crucial.}. 
  \begin{figure}[tbp]
    \centering
    \includegraphics[width=7.8cm]{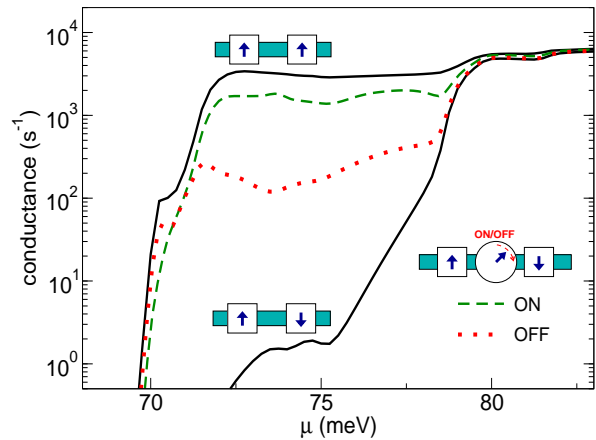}
    \caption{(Color online) Hybrid devices combining the SF and SR blocks described in Figs.~\ref{fig:SF} 
      and~\ref{fig:SR}, respectively.
      The full lines represent the total conductance for a 
      series of two SFs in ``open'' or ``closed'' configuration. 
      The dashed and dotted lines are for the total conductance of combined system made by the series of a SF($\uparrow$), 
      a SR(ON/OFF) and a SF($\downarrow$).
      The calculation has been performed for a gate-bias of $|\Delta|=150$~meV, with  
      $L_C=400 a_{0}$ for the SFs and $L_C=175 a_{0}$ for the SR.
    }
    \label{fig:SFET}
  \end{figure}

  Fig.~\ref{fig:SFET} shows the total conductance of a spintronic device made by 
  the composition of SFs and a SR as a function of $\mu$.
  The full black lines, as indicated, show the conductances of an ``open'' series, 
  i.e. the SF($\uparrow$)-SF($\uparrow$), and of a ``closed'' one SF($\uparrow$)-SF($\downarrow$).
  For $\mu$ in the operational region of the SF, the total conductance of the ``closed'' series is suppressed by almost 
  4 orders of magnitude with respect to the ``open'' one.  
  The remaining lines represent the conductance of a ``closed'' series including the SR described in Fig.~\ref{fig:SR}.
  The dashed green line---the SR(ON) case---is quite close to the conductance of the ``open'' series.
  The dotted red line---the SR(OFF) case---exhibits a conductance suppression by approximately a factor 
  $10$, with respect to the ON case.
  This expresses a measure for the efficiency of the SR, for which, in fact, we note that 
  $\chi_{\rm ON}/\chi_{\rm OFF}\approx 10$ for the corresponding data set (b) in the inset of Fig.~\ref{fig:SR}.

  \section{Discussion and conclusions}
  We now discuss the possibility of actually realizing the EPI in graphene devices.
  One of the few concrete examples of a FI is EuO.
  The first realization of the EPI coupling, originating a spin-splitting, has been experimentally 
  proven in a EuO/superconductor interface~\cite{tedrow1986}.
  The possibility to incorporate FI in nanostructures has been recognized to be extremely attractive for the realization 
  of spintronic nanodevices and recently much effort has been put into the development of the FI technology.
  In particular, important steps have been made in the control of the epitaxial 
  growth of EuO on Si and GaAs~\cite{schmehl2007,swartz2010}. 
  EuO seems promising with its semiconducting gap of about $0.7$~eV~\cite{steeneken2002,santos2008} and 
  the possibility to be grown in thin films of a few nm thickness~\cite{santos2008,muller2009}. 
  Regarding the practical realization of the device, a suitable FI should have  
  a sufficiently large bandgap and retain its properties when 
  grown in thin films.

  The occurrence of a EPI for graphene deposited on a FI has been 
  proposed by several authors~\cite{semenov2007, haugen2008, semenov2008}, with a tentative 
  estimation of the expected Zeeman splitting of $E_z\approx5$~meV~\cite{haugen2008}.
  When a BG is placed in contact with a FI, the EPI, being short-ranged, affects only the U graphene 
  layer, which is in direct contact with the FI~\cite{semenov2008}.
  In fact,  applying the contact exchange model between magnetic ions and itinerant electrons 
  proposed in Ref.~\onlinecite{merkulov1999}, and using the asymptotic atomic wave functions for 
  carbon~\cite{smirnov2001}, it is easy to show (see Appendix~\ref{app3}) 
  that the ratio between the EPI strength on the L and U layer is on the order of $e^{-2\kappa d_{0}}\approx10^{-3}$, 
  with $d_{0}=3.4$~\AA~the interlayer distance in BG and $\kappa\approx0.91a_B^{-1}$ the asymptotic 
  exponent for C~\cite{smirnov2001}.

  In summary, we have demonstrated that the exchange proximity interaction in bilayer graphene in contact 
  to a ferromagnetic insulator can be exploited as a means for electrical spin manipulation.
  We have shown that this system acts both as a switchable spin-filter or spin-rotator, which are basic 
  building blocks for spintronics.
  As an example, we have shown how to realize a complete spintronic structure for the creation, manipulation and 
  detection of spin currents---a spin FET, out of an initially unpolarized stream of electrons 
  and calculated its operational efficiency with a transfer matrix approach.

  \begin{acknowledgments}
    We thank the DFG for financial support via the Emmy Noether program.
  \end{acknowledgments}

  \appendix

  \section{Bilayer graphene eigenspinors\label{app1}}
  
  We consider the BG Hamiltonian
  \begin{equation}
    H =  \left(\begin{array}{cccc}
        -\frac{\Delta}{2}+h_m  & v_f k_+ &  t_\perp & 0 \\
        v_f k_- & -\frac{\Delta}{2}+h_m  & 0 & 0 \\
        t_\perp & 0 & \frac{\Delta}{2} & v_f k_- \\
        0 & 0 & v_f k_+ & \frac{\Delta}{2}
      \end{array}\right)
    \label{H}
  \end{equation}
  with $k_\pm= k_x \pm i k_y$ and where $\Delta$, $v_f$ and $t_\perp$ have been introduced in Section \ref{sec1}.
  EPI affects only the U plane and is contained in $h_m$ (Eq.~\ref{eq:EPI}), all other terms are proportional to the identity in the spin subspace. 
  The Hamiltonian acts on the spinor
  \begin{equation}
    \Psi = \left(
      \begin{array}{l}
        \chi_A \\ \chi_B \\ \chi_{B'} \\ \chi_{A'}
      \end{array}
    \right) \frac{e^{i k_x x} e^{i k_y y}}{\sqrt{L_x L_y}},
    \label{eq:wavefunction}
  \end{equation}
  where $A$, $B$ refer to the two inequivalent sublattices on the U
  BG layer, $A'$,$B'$ to that of the L one.
  $L_x$ and $L_y$ are the channel dimensions along $X$ and $Y$ directions.
  Now we distinguish the two spin components along the $Z$ axis,
  perpendicular to the plane, therefore $\chi_X$, with
  $X=A$, $B$, $A'$, $B'$, has to be regarded as a two-component spinor
  \begin{equation}
    \chi_X = \left(
      \begin{array}{l}
        \phi_{X\uparrow} \\ \phi_{X\downarrow}
      \end{array}
    \right).
  \end{equation}

  We introduce the following notation, similar to that chosen in Ref.~\onlinecite{barbier2009}, 
  \begin{equation}
    \begin{array}{ccc}
      \delta = \frac{V_U-V_L}{2 \hbar v_F} ~&~ \varepsilon = \frac{E}{\hbar v_F} ~&~   u_0=\frac{V_U+V_L}{2 \hbar v_F} \\
      &&\\
      \varepsilon'=\varepsilon-u_0 ~&~  t'=\frac{t_\perp}{\hbar v_F} ~&~  \alpha'=\frac{E_{Z}}{\hbar v_F}.
    \end{array}
    \label{eq:notation}
  \end{equation}

  \subsection{Spinors of BG without EPI}
  In this section, we give the analytical expressions for the spinors of normal BG, 
  i.e. without EPI and neglecting the trigonal warping effects.
  Spin is degenerate and therefore we can consider $\chi_{X}$ as scalar complex numbers.
  The Hamiltonian system $(H_0-E)\Psi=0$ leads to the secular equation
  \begin{eqnarray}
    \left[(\varepsilon'-\delta)^2-k^2\right]\left[(\varepsilon'+\delta)^2-k^2\right]-t'^2(\varepsilon'^2-\delta^2) = 0,
  \end{eqnarray}
  with $k^2=k_x^2+k_y^2$.
  If we solve for the energy we obtain the BG eigenstates 
  \begin{equation}
    (\varepsilon_{\pm}')^2 = \delta^2 + k^2+\frac{t'^2}{2}
    \pm t'\sqrt{\frac{t'^2}{4} + 4\frac{\delta^2k^2}{t'^2} + k^2}.
    \label{eq:energy_NOEI}
  \end{equation}
  If, instead, we solve for $k_x$ we obtain the BG modes consistent with energy $E$ and transverse wavevector $k_y$ 
  \begin{equation}
    k_x^2 = -k_y^2 + \varepsilon'^2+\delta^2 \pm t'\sqrt{\varepsilon'^2-\delta^2+4\frac{\varepsilon'^2\delta^2}{t'^2}}.
    \label{eq:kx_NOEI}
  \end{equation}
  
  The spinor components can be expressed (for $\varepsilon' \neq \pm \delta$) as 
  \begin{eqnarray}
    \chi_{A} &=& \frac{\mathcal{B}}{t' (\varepsilon' + \delta)} \chi_{B'}\nonumber\\
    \chi_{B} &=& \frac{\mathcal{B} (k_x-i k_y)}{t' (\varepsilon'^2 - \delta^2)} \chi_{B'}\nonumber\\
    \chi_{A'} &=& \frac{(k_x + ik_y)}{\varepsilon' + \delta} \chi_{B'}\nonumber\\
    \chi_{B'} &=& \frac{t' (\varepsilon'^2 - \delta^2)}{\sqrt{\mathcal{A}|\mathcal{B}|^2 + t'^2
        (\varepsilon' - \delta)^2 \mathcal{C}}},
  \end{eqnarray}
  where we adopted the following notation
  \begin{eqnarray*}
    \mathcal{A} &=& (\varepsilon' - \delta)^2 + |k_x-i k_y|^2 \\
    \mathcal{B} &=& (\varepsilon' + \delta)^2 - k^2 \\
    \mathcal{C} &=& (\varepsilon' + \delta)^2 + |k_x+i k_y|^2\\ 
    \mathcal{D} &=& (\varepsilon' - \delta)^2 - k^2.
  \end{eqnarray*}

  \subsubsection{Layer localization and trigonal warping corrections}
  The probability to find the electron, of a specific eigenfunction, on the U plane is given by
  \begin{equation}
    P^U = |\chi_A|^2 + |\chi_B|^2 = \frac{1}{1+ \frac{\mathcal{A}\mathcal{B}^2}{t'^2(\varepsilon'-\delta)^2 \mathcal{C}}},
  \end{equation}
  and correspondingly $P^L=1-P^U$.
  For our purpose, the most important part of the dispersion curve is the Mexican hat region ($|k|\approx k_{min}$).
  For this reason we now describe in more details this minimum for the first conduction band.
  Its wavevector, energy and layer projection are given by
  \begin{eqnarray}
    k_{min} &=& \delta \sqrt{1+X^2} \nonumber\\
    \varepsilon'_{min} &=& \pm \delta X \nonumber\\
    P^U_{min} &=& \frac{1}{1+ \frac{t'^2+4\delta^2}{4 \delta^2} \frac{1+X^2+X}{1+X^2-X} \left(X^2-1\right)},
    \label{eq:kmin}
  \end{eqnarray}
  with $X=\frac{t'}{\sqrt{4 \delta^2+t'^2}}$.
  \begin{figure}
    \centering
    \includegraphics[width=7.cm]{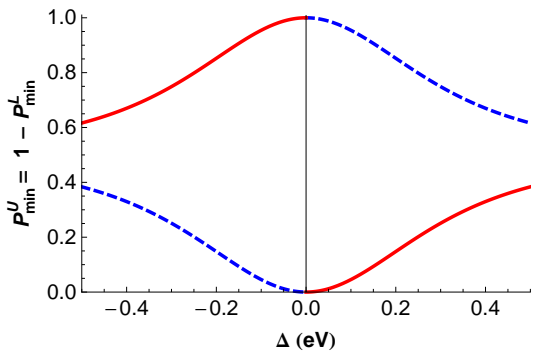}
    \caption{Spinor-projection on the U BG plane for the eigenstate corresponding to $k_{min}$ of the first conduction (full line) 
      and valence band (dashed line), as a function of the layer potential bias $\Delta$.}
    \label{fig:minimo}	
  \end{figure}
  In Fig.~\ref{fig:minimo}, we show the layer localization properties of the first conduction band 
  minimum with a full curve (and the first valence band maximum with a dashed line).
  For a sufficiently small $|\Delta|$, the corresponding states are strongly 
  localized on the upper or on the lower graphene plane, 
  depending on the sign of the applied bias $\Delta$.
  Similarly, the low energy states of the first conduction band share analogously strong layer localization properties
  with the conduction band minimum.
  In particular, the layer localization complete for the $k=0$ state ($P^{U}=1$ or $0$).

\begin{figure}[tb]
    \centering
    \includegraphics[width=7.cm]{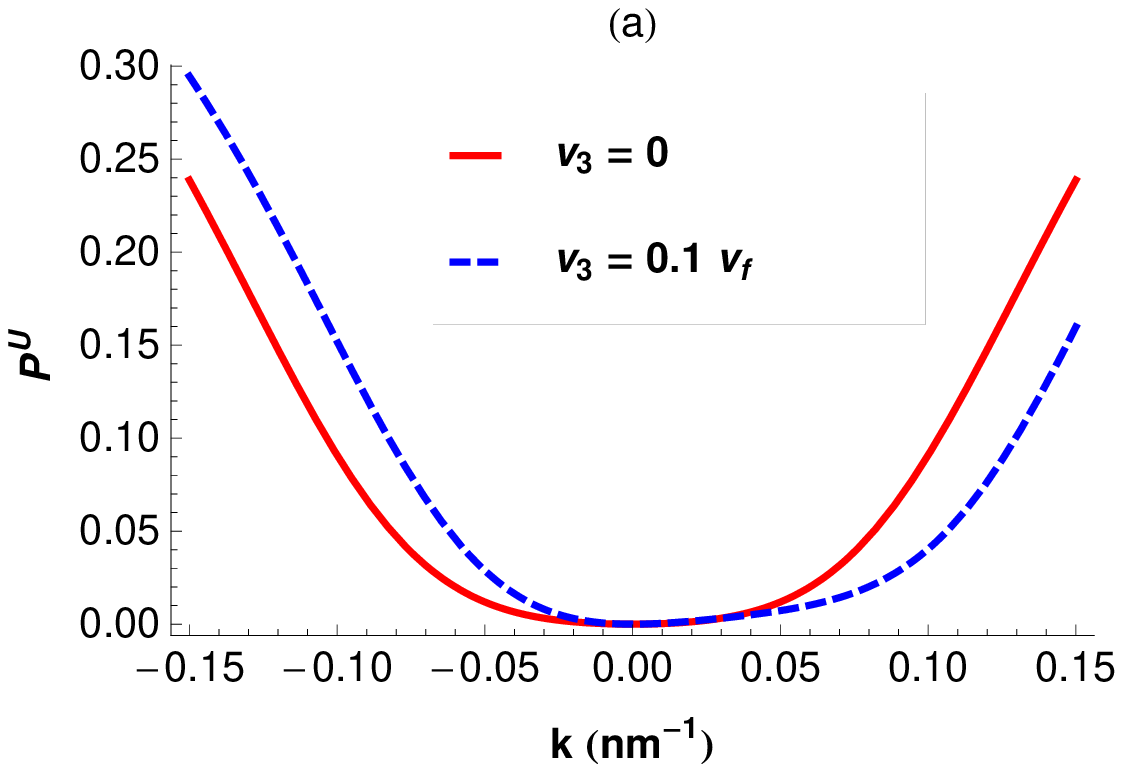}  \includegraphics[width=7.cm]{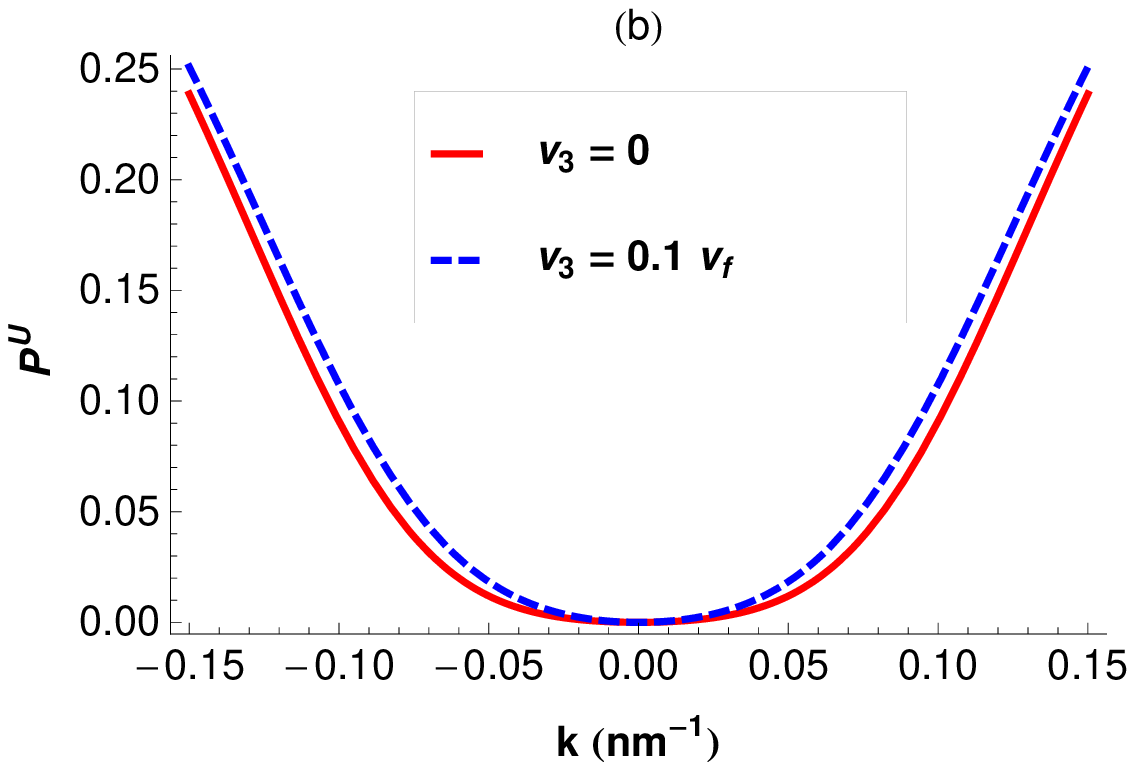}
    \caption{Projection of the eigenstates of the first conduction band on the U BG plane, with (dashed line) 
      and without (full line) trigonal warping corrections, for an applied bias of $\Delta=40$~meV.
      (a) is along the $X$ axis and (b) along the $Y$ axis.}
    \label{fig:trig}
  \end{figure}

  We analyze, now, the effect of the trigonal warping correction~\cite{nilsson2008} 
  on the layer localization properties of BG.
  The trigonal warping correction, acting on the spinor in Eq.~\ref{eq:wavefunction}, is  
  \begin{equation}
    H_3 = v_3 \frac{\sigma_0-\sigma_z}{2} (k_x\tau_x-k_y\tau_y),
  \end{equation}
  with $\sigma$, $\tau$ the sublattice and layer Pauli matrices, respectively. 
  We have solved the system with a finite trigonal warping with $v_3/v_f=0.1$~\cite{nilsson2008}, 
  found its eigenstates and calculated their layer projections.
  In Fig.~\ref{fig:trig}, we compare the projection on the $U$ plane for the system without 
  trigonal warping (solid curve) and including the trigonal correction (dashed curve) 
  for the dispersion along the $X$ axis in (a) and $Y$ axes in (b) for $\Delta=40$~meV, 
  energy for which the isotropic Mexican hat dispersion is heavily distorted to a trigonal symmetry.
  In both cases, the low energy states are essentially localized on the L layer.
  We conclude that the layer projection properties are only slightly affected by the presence of the trigonal 
  warping in the BG Hamiltonian and the mechanism which permit to turn ON or OFF the BG spintronic 
  functionalities is still valid.

  The presence of the trigonal warping can however effect the transmission of the system.
  The spin filtering properties are unaffected, as long as a gap is present with spin-splitted bands.
  The spin rotation effect is due to the interference effect between different spin components propagating 
  with different wavevectors from the Mexican hat spectral region.
  The trigonal warping distortion of the Mexican hat can change the accumulated phase-difference and 
  therefore leads to differences in the form of the spin-flip transmission resonance.
  The distortion is more pronounced at smaller applied gate bias $|\Delta|$~\cite{nilsson2008}, inducing a 
  progressive angle dependence of the spin-rotating properties.
  In principle, also in this small-gap regime, spin rotation could still be exploited 
  in angle-selective transport experiments.
  Instead, for sufficiently large gate bias (and therefore semiconducting gap), the trigonal warping effect is limited to 
  a minor distortion of the Mexican hat and the properties of the system are essentially unchanged.
  This scenario is met for $|\Delta|=150$~meV used in the paper.

  \subsection{Spinors of BG with EPI}
  In this section we derive the analytical expressions for the BG spinors in 
  the case of a finite EPI acting on the U plane.
  EPI is not diagonal in the spin components and the spin variables are individually addressed.
  Still the Hamiltonian system $(H-E)\Psi=0$, can be analytically solved leading to the secular equation
  {\small\begin{eqnarray*}
    \left[\vspace{-0.1cm} t'^2 \vspace{-0.1cm} (\varepsilon'^2 - \delta^2) \vspace{-0.1cm} - \vspace{-0.1cm} \mathcal{B}\left(\mathcal{D}+\frac{\alpha'^2}{4}\right) \right]^2 
    \vspace{-0.1cm} = \vspace{-0.1cm}\frac{\alpha'^2}{4}\vspace{-0.05cm}\left[2\mathcal{B}(\varepsilon'-\delta) -t'^2 (\varepsilon'+\delta)\right]^2.
    \label{eq:characteristics_M}
  \end{eqnarray*}}
  The secular equation solved for $k_x$ has in general $8$ complex solutions
  {\small\begin{equation}
    k_x^2 = -P_1 
    \left(
      \begin{array}{l} 
        -\\-\\+\\+ 
      \end{array}
    \right) 
    P_2 
    \left(
      \begin{array}{l} 
        -\\+\\-\\+ 
      \end{array}\right) 
    \sqrt{
      \left[
        P1
        \left(
          \begin{array}{l} 
            +\\+\\-\\- 
          \end{array}
        \right) 
        P_2 
      \right]^2 
      - P_3 
      \left(
        \begin{array}{l} 
          -\\-\\+\\+ 
        \end{array}
      \right) 
      P_4 
    },
    \label{eq:modes_M}
  \end{equation}}
  with 
  \begin{eqnarray*}
    P_1 &=& \frac{1}{2} \left[\tilde{\mathcal{B}}-\tilde{\mathcal{D}}-\frac{\alpha'^2}{4}\right]\\
    P_2 &=& \frac{\alpha'}{2} (\varepsilon'^2-\delta^2)\\
    P_3 &=& -\tilde{\mathcal{B}}\tilde{\mathcal{D}} - t'^2 (\varepsilon'^2-\delta^2)-\frac{\alpha'^2}{4}\tilde{\mathcal{B}}\\
    P_4 &=& \frac{\alpha'}{2} \left[2\tilde{\mathcal{B}}(\varepsilon'-\delta) - t'^2 (\varepsilon'+\delta)\right],
  \end{eqnarray*}
  where $\tilde{\mathcal{B}}= \mathcal{B}+k_x^2$ and $\tilde{\mathcal{D}}=\mathcal{D}+k_x^2$.
  Real solutions correspond to propagating modes in the region with EPI interaction while modes with an 
  finite imaginary part give exponentially decaying modes at the border of the EPI region.
  
  We omit the the expressions for the spinor components.
  We obtained them with straightforward derivation, from the secular equation, as a function of $\phi_{B'\downarrow}$, which  
  in a second time we numerically fixed using the normalization condition.

  \section{Multiple barriers: transfer matrix method \label{app2}}
  Let us consider a 1D channel with a finite number of modes $M$ in which a scattering region is present.
  For each scattering center, we can fictitiously divide the system in a left-side (LS) and a right-side (RS) leads, 
  which we assume to be semi-infinite.
  The wavefunctions in the LS and RS leads are described by
  \begin{eqnarray}
    \psi_L &=& \sum_a \left( I^{(L)}_a \phi^{(+)}_a + O^{(L)}_a \phi^{(-)}_a \right)\\
    \psi_R &=& \sum_a \left( I^{(R)}_a \phi^{(-)}_a + O^{(R)}_a \phi^{(+)}_a \right),
  \end{eqnarray}   
  where $\phi^{(\pm)}_a$ is the $a$-th mode of the channel carrying an unity of current,
  where $\pm$ stands for forward-going (+), i.e. from the LS to the RS, and backward-going (-).
  $\vec{I}^{(L,R)}$ and $\vec{O}^{(L,R)}$ are the coefficient vectors, in the $LS$ and $RS$ regions, 
  for modes which are incoming towards the scattering center and outgoing from it, respectively.
  
  We define the scattering matrix $\mathbf{S}$ and the transfer matrix $\mathbf{T}$ through the following relations
  \begin{eqnarray}
    \left(
      \begin{array}{l}
        \vec{O}^{(R)}\\
        \vec{O}^{(L)}
      \end{array}
    \right)
    = \mathbf{S}
    \left(
      \begin{array}{l}
        \vec{I}^{(L)}\\
        \vec{I}^{(R)}
      \end{array}
    \right)\nonumber\\
    \left(
      \begin{array}{l}
        \vec{O}^{(R)}\\
        \vec{I}^{(R)}
      \end{array}
    \right)
    = \mathbf{T}
    \left(
      \begin{array}{l}
        \vec{I}^{(L)}\\
        \vec{O}^{(L)}
      \end{array}
    \right).
    \label{eq:defST}
  \end{eqnarray}
  It is immediate to identify the elements of the scattering matrix with reflection and transmission coefficients, so that
  \begin{equation}
    \mathbf{S}=
    \left(
      \begin{array}{ll}
        \mathbf{t} & \mathbf{r'}\\
        \mathbf{r} & \mathbf{t'}
      \end{array}
    \right),
  \end{equation}
  but we are interested in the transfer matrix because it is \emph{multiplicative}, i.e. the transfer matrix of a 
  series of scatterers is given by the \emph{ordered multiplication} 
  of the individual transfer matrices, for each one of the scattering centers, or
  \begin{equation}
    \mathbf{T}_{tot} = \prod_{n=1}^{n=N} \mathbf{T}_n. 
  \end{equation}
  We can obtain an expression for the transfer matrix in terms of reflection and transmission 
  coefficients by comparing the action of the $\mathbf{S}$ and  $\mathbf{T}$ matrices in Eq.~\ref{eq:defST}:
  \begin{eqnarray}
    \vec{O}^{(R)} &=&  \mathbf{t} \vec{I}^{(L)} + \mathbf{r'}  \vec{I}^{(R)}\nonumber\\
    \vec{O}^{(L)} &=&  \mathbf{t'} \vec{I}^{(R)} + \mathbf{r}  \vec{I}^{(L)}\nonumber\\
    \vec{O}^{(R)} &=&  \mathbf{T}_{1,1} \vec{I}^{(L)} +  \mathbf{T}_{1,2} \vec{O}^{(L)}\nonumber\\
    \vec{I}^{(R)} &=&  \mathbf{T}_{2,2} \vec{O}^{(L)} +  \mathbf{T}_{2,1} \vec{I}^{(L)}.
  \end{eqnarray}
  We obtain
  \begin{eqnarray}
    \mathbf{T}_{2,2} &=& \left[\mathbf{t'}\right]^{-1}\nonumber\\
    \mathbf{T}_{2,1} &=& -\left[\mathbf{t'}\right]^{-1}\mathbf{r}\nonumber\\
    \mathbf{T}_{1,1} &=& \mathbf{t} - \mathbf{r'}\left[\mathbf{t'}\right]^{-1}\mathbf{r}\nonumber\\
    \mathbf{T}_{1,2} &=& \mathbf{r'} \left[\mathbf{t'}\right]^{-1}.
  \end{eqnarray}
  In practice, in order to calculate the transfer matrix for each one of the scatterers, the transmission and reflection 
  coefficients for a particle approaching from the $LS$ and $RS$ of the scattering center are needed.

  From the total transfer matrix, it is then possible to obtain the transmission and the reflection properties 
  of the overall system by the following relations 
  \begin{eqnarray}
    \mathbf{t'} &=& \left[\mathbf{T}_{2,2}\right]^{-1}\nonumber\\
    \mathbf{r} &=& -\left[\mathbf{T}_{2,2}\right]^{-1} \mathbf{T}_{2,1}\nonumber\\
    \mathbf{t} &=& \mathbf{T}_{1,1} - \mathbf{T}_{1,2}\left[\mathbf{T}_{2,2}\right]^{-1} \mathbf{T}_{2,1}\nonumber\\
    \mathbf{r'} &=& \mathbf{T}_{1,2} \left[\mathbf{T}_{2,2}\right]^{-1}.
  \end{eqnarray}

  \subsection{Properties of S and T matrices}
  The scattering matrix has to be unitary in order to ensure charge conservation in a barrier, i.e.  $I_{in}=I_{out}$.
  Explicitly
  \begin{eqnarray*}
    \left|
      \left(
        \begin{array}{l}
          \vec{I}^{(R)}\\
          \vec{I}^{(L)}
        \end{array}
      \right)
    \right|^2=
    \left|
      \left(
        \begin{array}{l}
          \vec{O}^{(R)}\\
          \vec{O}^{(L)}
        \end{array}
      \right)
    \right|^2
    =
    \left(
      \begin{array}{l}
        \vec{I}^{(R)}\\
        \vec{I}^{(L)}
      \end{array}
    \right)^\dag
    \mathbf{S}^\dag
    \mathbf{S}
    \left(
      \begin{array}{l}
        \vec{I}^{(R)}\\
        \vec{I}^{(L)}
      \end{array}
    \right),
  \end{eqnarray*}
  which is satisfied if $\mathbf{S}^\dag \mathbf{S}=1$ and therefore $\mathbf{S}^\dag=[\mathbf{S}]^{-1}$.
  A well-known consequence of the unitarity of the scattering matrix is that $|\mathbf{t}|=|\mathbf{t'}|$ 
  and $|\mathbf{r}|=|\mathbf{r'}|$, valid for any kind of elastic scatterer.

  The condition for a stationary equilibrium current through the barrier is given by  
  \begin{equation}
    I=\left|\vec{I}^{(L)}\right|^2 -\left|\vec{O}^{(L)}\right|^2 = \left|\vec{I}^{(R)}\right|^2 -\left| \vec{O}^{(R)}\right|^2,
  \end{equation}
  which imposes the following property on the transfer matrix $\mathbf{T}^\dag \mathbf{\sigma}_z  \mathbf{T} =\mathbf{\sigma}_z$.

  \section{Estimate of EPI effects on the distant BG layer\label{app3}}
  The goal of this section is to provide an estimation of the relative importance of the EPI of the two layers of BG, 
  when the U layer is placed in direct contact to the FI surface.
  We will consider the graphene layers and the FI surface oriented along the $XY$ plane with the FI surface at $z=0$, 
  the U plane centered at $z=L_U$ and the L plane centered at $z=L_L$, with $L_L-L_U=d_0$ the BG interlayer distance.  
  As noted in Ref.~\onlinecite{tokuyasu1988}, the exchange coupling between an itinerant electron and the 
  local moments in FIs (like EuO) typically dominates the coupling to the magnetization.
  The exchange potential for a mobile electron, arising because of the exchange interaction with 
  core electrons in a magnetic ion, is modeled by~\cite{merkulov1999} 
  \begin{equation}
    V_{ex}(\vec{r}) = -j \sum_i^{ions} \delta(|\vec{r}-\vec{R}_i|)\hat{S}_i \hat{S},
    \label{eq:ex2}
  \end{equation}  
  where $\hat{S}_i$ is the ion's total spin, $\hat{S}$ is the electron spin operator and $j$ an exchange energy parameter.  
  This expression is used in Ref.~\onlinecite{merkulov1999} to model the exchange potential of conduction 
  electrons due to the presence of localized core $d$-electrons in Mn magnetic ions.
  The assumption is that the wavefunctions of mobile electrons can be considered approximatively 
  constant in the range of variation of the magnetic ion's occupied orbitals.

  In our model the mobile electrons are the graphene bilayer conduction and valence band electrons.
  In the direction perpendicular to the graphene plane, this material is practically one-atom thick 
  and its  conduction and valence bands can essentially be described by the carbon atoms $p_z$-orbital.
  The tails of the graphene $p_z$ orbital enter the FI, where the magnetic ions are distributed.
  We assume an homogeneous distribution of magnetic ions inside the FI, with density $n_{ions}$ 
  in the FI with fixed average spin polarization $\langle S_{i} \rangle$ along $Z$.
  We obtain the following exchange potential for the graphene electrons 
  \begin{equation}
    V_{ex}(\vec{r}) = -j n_{ions} \Theta(z-L) \hat{S}_z \langle\hat{S}_i\rangle,
    \label{eq:ex}
  \end{equation} 
  where $\Theta(x)$ is the step function and $z=L$ identify the surface of the FI.

  An established result in atomic physics~\cite{smirnov2001} is that we can describe the asymptotic behavior of wave functions
  for valence electrons in an atom, at large distances, as $\psi(\vec{r})= R_{n,m}(r) Y_{m,l}$, with $R(r)$
  \begin{equation}
    R(r) = A r^{1/\kappa-1} e^{-r\kappa},
    \label{eq:R}
  \end{equation} 
  with $\kappa=\sqrt{2 I}$, where $I$ is the ionization potential for that electron in the atom.
  A $p_z$-electron on a graphene plane will therefore be described as
  \begin{equation}
    \psi(r,\theta) \propto  r^{1/\kappa-1} e^{-r\kappa} \cos{\theta}.
    \label{eq:psi}
  \end{equation}

  The direct EPI energy between the FI and a graphene $\pi$-orbital from the U graphene plane is proportional to
  \small{\begin{eqnarray}
    E_{Z}^{(U)} &\propto& \int_{L-L_1}^{+\infty} dr~r^{2/\kappa} e^{-2r\kappa} \int_0^{\arccos{\frac{L-L_U}{r}}} d\theta \cos^2{\theta}=\nonumber\\
    &=& \frac{(L-L_U)^{2/\kappa+1}}{2} \mathcal{I}_1, 
    \label{eq:J1}
  \end{eqnarray}} 
  with $t=r/(L-L_U)$ and
  \small{
    \begin{eqnarray}
      \mathcal{I}_1 =\int_{1}^{+\infty} dt~t^{2/\kappa} e^{-2t (L-L_U)\kappa} 
    \left[\frac{\sqrt{t^2-1}}{t^2}+\arccos{\frac{1}{t}}\right].
  \end{eqnarray}} 
  We now calculate the ratio $E_Z^{(L)}/E_Z^{(U)}$, where $E_Z^{(L)}$ is the magnitude of the EPI with a carbon $p_z$ orbital 
  from the $L$ graphene plane, which is further away from the FI surface, than the $U$ one. 
  Observing the form of the EPI in Eq.~\ref{eq:J1} and using the fact that $L-L_L>L-L_U$, 
  we can infer the following condition for $E_Z^{(L)}$ 
  \begin{eqnarray}
    E_Z^{(L)} < \frac{(L-L_L)^{2/\kappa+1}}{2} e^{-2(L_U-L_L)\kappa}   \mathcal{I}_1,
    \label{eq:J2}
  \end{eqnarray} 
  and therefore the ratio
  \begin{eqnarray}
    \frac{E_Z^{(L)}}{E_Z^{(U)}} <  e^{-2 d_0 \kappa}. 
    \label{eq:ratio}
  \end{eqnarray}  
  For two neighboring graphene layers (interlayer distance $d_0$ around $0.34$~nm), 
  and employing the value for $C$ atoms $\kappa=0.910$~${\rm a}_{\rm B}^{-1}$ 
  from Ref.~\onlinecite{smirnov2001}, we conclude that the ratio of the exchange 
  interaction ($J_L/J_U$) is of the order of $2 \times 10^{-3}$. 
  We can therefore safely neglect the EPI effect on the lower layer.

  \bibliography{bibf}

\begin{thebibliography}{35}
\expandafter\ifx\csname natexlab\endcsname\relax\def\natexlab#1{#1}\fi
\expandafter\ifx\csname bibnamefont\endcsname\relax
  \def\bibnamefont#1{#1}\fi
\expandafter\ifx\csname bibfnamefont\endcsname\relax
  \def\bibfnamefont#1{#1}\fi
\expandafter\ifx\csname citenamefont\endcsname\relax
  \def\citenamefont#1{#1}\fi
\expandafter\ifx\csname url\endcsname\relax
  \def\url#1{\texttt{#1}}\fi
\expandafter\ifx\csname urlprefix\endcsname\relax\def\urlprefix{URL }\fi
\providecommand{\bibinfo}[2]{#2}
\providecommand{\eprint}[2][]{\url{#2}}

\bibitem[{\citenamefont{Geim and Novoselov}(2007)}]{geim2007}
\bibinfo{author}{\bibfnamefont{A.~K.} \bibnamefont{Geim}} \bibnamefont{and}
  \bibinfo{author}{\bibfnamefont{K.~S.} \bibnamefont{Novoselov}},
  \bibinfo{journal}{Nat. Mat.} \textbf{\bibinfo{volume}{6}},
  \bibinfo{pages}{183} (\bibinfo{year}{2007}).

\bibitem[{\citenamefont{Tombros et~al.}(2007)\citenamefont{Tombros, Jozsa,
  Popinciuc, Jonkman, and van Wees}}]{tombros2007}
\bibinfo{author}{\bibfnamefont{N.}~\bibnamefont{Tombros}},
  \bibinfo{author}{\bibfnamefont{C.}~\bibnamefont{Jozsa}},
  \bibinfo{author}{\bibfnamefont{M.}~\bibnamefont{Popinciuc}},
  \bibinfo{author}{\bibfnamefont{H.~T.} \bibnamefont{Jonkman}},
  \bibnamefont{and} \bibinfo{author}{\bibfnamefont{B.}~\bibnamefont{van Wees}},
  \bibinfo{journal}{Nature} \textbf{\bibinfo{volume}{448}},
  \bibinfo{pages}{571} (\bibinfo{year}{2007}).

\bibitem[{\citenamefont{Tombros et~al.}(2008)\citenamefont{Tombros, Tanabe,
  Veligura, Jozsa, Popinciuc, Jonkman, and van Wees}}]{tombros2008}
\bibinfo{author}{\bibfnamefont{N.}~\bibnamefont{Tombros}},
  \bibinfo{author}{\bibfnamefont{S.}~\bibnamefont{Tanabe}},
  \bibinfo{author}{\bibfnamefont{A.}~\bibnamefont{Veligura}},
  \bibinfo{author}{\bibfnamefont{C.}~\bibnamefont{Jozsa}},
  \bibinfo{author}{\bibfnamefont{M.}~\bibnamefont{Popinciuc}},
  \bibinfo{author}{\bibfnamefont{H.~T.} \bibnamefont{Jonkman}},
  \bibnamefont{and} \bibinfo{author}{\bibfnamefont{B.~J.} \bibnamefont{van
  Wees}}, \bibinfo{journal}{Phys. Rev. Lett.} \textbf{\bibinfo{volume}{101}},
  \bibinfo{pages}{046601} (\bibinfo{year}{2008}).

\bibitem[{\citenamefont{Ertler et~al.}(2009)\citenamefont{Ertler, Konschuh,
  Gmitra, and Fabian}}]{ertler2009}
\bibinfo{author}{\bibfnamefont{C.}~\bibnamefont{Ertler}},
  \bibinfo{author}{\bibfnamefont{S.}~\bibnamefont{Konschuh}},
  \bibinfo{author}{\bibfnamefont{M.}~\bibnamefont{Gmitra}}, \bibnamefont{and}
  \bibinfo{author}{\bibfnamefont{J.}~\bibnamefont{Fabian}},
  \bibinfo{journal}{Phys. Rev. B} \textbf{\bibinfo{volume}{80}},
  \bibinfo{pages}{041405} (\bibinfo{year}{2009}).

\bibitem[{\citenamefont{Han and Kawakami}()}]{han2011}
\bibinfo{author}{\bibfnamefont{W.}~\bibnamefont{Han}} \bibnamefont{and}
  \bibinfo{author}{\bibfnamefont{R.~K.} \bibnamefont{Kawakami}},
  \bibinfo{note}{arXiv:1012.3435v1}.

\bibitem[{\citenamefont{Yang et~al.}()\citenamefont{Yang, Balakrishnan, Volmer,
  Avsar, Jaiswal, Samm, Ali, Pachoud, Zeng, Popinciuc et~al.}}]{yang2011}
\bibinfo{author}{\bibfnamefont{T.~Y.} \bibnamefont{Yang}},
  \bibinfo{author}{\bibfnamefont{J.}~\bibnamefont{Balakrishnan}},
  \bibinfo{author}{\bibfnamefont{F.}~\bibnamefont{Volmer}},
  \bibinfo{author}{\bibfnamefont{A.}~\bibnamefont{Avsar}},
  \bibinfo{author}{\bibfnamefont{M.}~\bibnamefont{Jaiswal}},
  \bibinfo{author}{\bibfnamefont{J.}~\bibnamefont{Samm}},
  \bibinfo{author}{\bibfnamefont{S.~R.} \bibnamefont{Ali}},
  \bibinfo{author}{\bibfnamefont{A.}~\bibnamefont{Pachoud}},
  \bibinfo{author}{\bibfnamefont{M.}~\bibnamefont{Zeng}},
  \bibinfo{author}{\bibfnamefont{M.}~\bibnamefont{Popinciuc}},
  \bibnamefont{et~al.}, \bibinfo{note}{arXiv:1012.1156v1}.

\bibitem[{\citenamefont{Han et~al.}(2010)\citenamefont{Han, Pi, McCreary, Li,
  Wong, Swartz, and Kawakami}}]{han2010}
\bibinfo{author}{\bibfnamefont{W.}~\bibnamefont{Han}},
  \bibinfo{author}{\bibfnamefont{K.}~\bibnamefont{Pi}},
  \bibinfo{author}{\bibfnamefont{K.~M.} \bibnamefont{McCreary}},
  \bibinfo{author}{\bibfnamefont{Y.}~\bibnamefont{Li}},
  \bibinfo{author}{\bibfnamefont{J.~J.~I.} \bibnamefont{Wong}},
  \bibinfo{author}{\bibfnamefont{A.~G.} \bibnamefont{Swartz}},
  \bibnamefont{and} \bibinfo{author}{\bibfnamefont{R.~K.}
  \bibnamefont{Kawakami}}, \bibinfo{journal}{Phys. Rev. Lett.}
  \textbf{\bibinfo{volume}{105}}, \bibinfo{pages}{167202}
  (\bibinfo{year}{2010}).

\bibitem[{\citenamefont{Trauzettel et~al.}(2007)\citenamefont{Trauzettel,
  Bulaev, Loss, and Burkard}}]{trauzettel2007}
\bibinfo{author}{\bibfnamefont{B.}~\bibnamefont{Trauzettel}},
  \bibinfo{author}{\bibfnamefont{D.~V.} \bibnamefont{Bulaev}},
  \bibinfo{author}{\bibfnamefont{D.}~\bibnamefont{Loss}}, \bibnamefont{and}
  \bibinfo{author}{\bibfnamefont{G.}~\bibnamefont{Burkard}},
  \bibinfo{journal}{Nat. Phys.} \textbf{\bibinfo{volume}{3}},
  \bibinfo{pages}{192} (\bibinfo{year}{2007}).

\bibitem[{\citenamefont{Recher and Trauzettel}(2010)}]{recher2010}
\bibinfo{author}{\bibfnamefont{P.}~\bibnamefont{Recher}} \bibnamefont{and}
  \bibinfo{author}{\bibfnamefont{B.}~\bibnamefont{Trauzettel}},
  \bibinfo{journal}{Nanotechnology} \textbf{\bibinfo{volume}{21}},
  \bibinfo{pages}{302001} (\bibinfo{year}{2010}).

\bibitem[{\citenamefont{Yazyev}(2008)}]{yazyev2008}
\bibinfo{author}{\bibfnamefont{O.~V.} \bibnamefont{Yazyev}},
  \bibinfo{journal}{Nano Lett.} \textbf{\bibinfo{volume}{8}},
  \bibinfo{pages}{1011} (\bibinfo{year}{2008}).

\bibitem[{\citenamefont{Fischer et~al.}(2009)\citenamefont{Fischer, Trauzettel,
  and Loss}}]{fischer2009}
\bibinfo{author}{\bibfnamefont{J.}~\bibnamefont{Fischer}},
  \bibinfo{author}{\bibfnamefont{B.}~\bibnamefont{Trauzettel}},
  \bibnamefont{and} \bibinfo{author}{\bibfnamefont{D.}~\bibnamefont{Loss}},
  \bibinfo{journal}{Phys. Rev. B} \textbf{\bibinfo{volume}{80}},
  \bibinfo{pages}{155401} (\bibinfo{year}{2009}).

\bibitem[{\citenamefont{Min et~al.}(2006)\citenamefont{Min, Hill, Sinitsyn,
  Sahu, Kleinman, and MacDonald}}]{min2006}
\bibinfo{author}{\bibfnamefont{H.}~\bibnamefont{Min}},
  \bibinfo{author}{\bibfnamefont{J.~E.} \bibnamefont{Hill}},
  \bibinfo{author}{\bibfnamefont{N.~A.} \bibnamefont{Sinitsyn}},
  \bibinfo{author}{\bibfnamefont{B.~R.} \bibnamefont{Sahu}},
  \bibinfo{author}{\bibfnamefont{L.}~\bibnamefont{Kleinman}}, \bibnamefont{and}
  \bibinfo{author}{\bibfnamefont{A.~H.} \bibnamefont{MacDonald}},
  \bibinfo{journal}{Phys. Rev. B} \textbf{\bibinfo{volume}{74}},
  \bibinfo{pages}{165310} (\bibinfo{year}{2006}).

\bibitem[{\citenamefont{Huertas-Hernando
  et~al.}(2006)\citenamefont{Huertas-Hernando, Guinea, and
  Brataas}}]{huertas2006}
\bibinfo{author}{\bibfnamefont{D.}~\bibnamefont{Huertas-Hernando}},
  \bibinfo{author}{\bibfnamefont{F.}~\bibnamefont{Guinea}}, \bibnamefont{and}
  \bibinfo{author}{\bibfnamefont{A.}~\bibnamefont{Brataas}},
  \bibinfo{journal}{Phys. Rev. B} \textbf{\bibinfo{volume}{74}},
  \bibinfo{pages}{155426} (\bibinfo{year}{2006}).

\bibitem[{\citenamefont{Gmitra et~al.}(2009)\citenamefont{Gmitra, Konschuh,
  Ertler, Ambrosch-Draxl, and Fabian}}]{gmitra2009}
\bibinfo{author}{\bibfnamefont{M.}~\bibnamefont{Gmitra}},
  \bibinfo{author}{\bibfnamefont{S.}~\bibnamefont{Konschuh}},
  \bibinfo{author}{\bibfnamefont{C.}~\bibnamefont{Ertler}},
  \bibinfo{author}{\bibfnamefont{C.}~\bibnamefont{Ambrosch-Draxl}},
  \bibnamefont{and} \bibinfo{author}{\bibfnamefont{J.}~\bibnamefont{Fabian}},
  \bibinfo{journal}{Phys. Rev. B} \textbf{\bibinfo{volume}{80}},
  \bibinfo{pages}{235431} (\bibinfo{year}{2009}).

\bibitem[{\citenamefont{Datta and Das}(1990)}]{datta1990}
\bibinfo{author}{\bibfnamefont{S.}~\bibnamefont{Datta}} \bibnamefont{and}
  \bibinfo{author}{\bibfnamefont{B.}~\bibnamefont{Das}},
  \bibinfo{journal}{Appl. Phys Lett.} \textbf{\bibinfo{volume}{56}},
  \bibinfo{pages}{665} (\bibinfo{year}{1990}).

\bibitem[{\citenamefont{Semenov et~al.}(2007)\citenamefont{Semenov, Kim, and
  Zavada}}]{semenov2007}
\bibinfo{author}{\bibfnamefont{Y.~G.} \bibnamefont{Semenov}},
  \bibinfo{author}{\bibfnamefont{K.~W.} \bibnamefont{Kim}}, \bibnamefont{and}
  \bibinfo{author}{\bibfnamefont{J.~M.} \bibnamefont{Zavada}},
  \bibinfo{journal}{Appl. Phys. Lett.} \textbf{\bibinfo{volume}{91}},
  \bibinfo{pages}{153105} (\bibinfo{year}{2007}).

\bibitem[{\citenamefont{Haugen et~al.}(2008)\citenamefont{Haugen,
  Huertas-Hernando, and Brataas}}]{haugen2008}
\bibinfo{author}{\bibfnamefont{H.}~\bibnamefont{Haugen}},
  \bibinfo{author}{\bibfnamefont{D.}~\bibnamefont{Huertas-Hernando}},
  \bibnamefont{and} \bibinfo{author}{\bibfnamefont{A.}~\bibnamefont{Brataas}},
  \bibinfo{journal}{Phys. Rev. B} \textbf{\bibinfo{volume}{77}},
  \bibinfo{pages}{115406} (\bibinfo{year}{2008}).

\bibitem[{\citenamefont{Semenov et~al.}(2008)\citenamefont{Semenov, Zavada, and
  Kim}}]{semenov2008}
\bibinfo{author}{\bibfnamefont{Y.~G.} \bibnamefont{Semenov}},
  \bibinfo{author}{\bibfnamefont{J.~M.} \bibnamefont{Zavada}},
  \bibnamefont{and} \bibinfo{author}{\bibfnamefont{K.~W.} \bibnamefont{Kim}},
  \bibinfo{journal}{Phys. Rev. B} \textbf{\bibinfo{volume}{77}},
  \bibinfo{pages}{235415} (\bibinfo{year}{2008}).

\bibitem[{\citenamefont{Castro et~al.}(2007)\citenamefont{Castro, Novoselov,
  Morozov, Peres, Lopes~dos Santos, Nilsson, Guinea, Geim, and
  Castro~Neto}}]{castro2007}
\bibinfo{author}{\bibfnamefont{E.~V.} \bibnamefont{Castro}},
  \bibinfo{author}{\bibfnamefont{K.~S.} \bibnamefont{Novoselov}},
  \bibinfo{author}{\bibfnamefont{S.~V.} \bibnamefont{Morozov}},
  \bibinfo{author}{\bibfnamefont{N.~M.~R.} \bibnamefont{Peres}},
  \bibinfo{author}{\bibfnamefont{J.~M.~B.} \bibnamefont{Lopes~dos Santos}},
  \bibinfo{author}{\bibfnamefont{J.}~\bibnamefont{Nilsson}},
  \bibinfo{author}{\bibfnamefont{F.}~\bibnamefont{Guinea}},
  \bibinfo{author}{\bibfnamefont{A.~K.} \bibnamefont{Geim}}, \bibnamefont{and}
  \bibinfo{author}{\bibfnamefont{A.~H.} \bibnamefont{Castro~Neto}},
  \bibinfo{journal}{Phys. Rev. Lett} \textbf{\bibinfo{volume}{99}},
  \bibinfo{pages}{216802} (\bibinfo{year}{2007}).

\bibitem[{\citenamefont{Oostinga et~al.}(2008)\citenamefont{Oostinga, Heersche,
  Liu, Morpurgo, and Vandersypen}}]{Oostinga2008}
\bibinfo{author}{\bibfnamefont{J.~B.} \bibnamefont{Oostinga}},
  \bibinfo{author}{\bibfnamefont{H.~B.} \bibnamefont{Heersche}},
  \bibinfo{author}{\bibfnamefont{X.}~\bibnamefont{Liu}},
  \bibinfo{author}{\bibfnamefont{A.~F.} \bibnamefont{Morpurgo}},
  \bibnamefont{and} \bibinfo{author}{\bibfnamefont{L.~M.~K.}
  \bibnamefont{Vandersypen}}, \bibinfo{journal}{Nature Mater.}
  \textbf{\bibinfo{volume}{7}}, \bibinfo{pages}{151} (\bibinfo{year}{2008}).

\bibitem[{\citenamefont{Zhang et~al.}(2009)\citenamefont{Zhang, Tang, Girit,
  Hao, Martin, Zettl, Crommie, Shen, and Wang}}]{zhang2009}
\bibinfo{author}{\bibfnamefont{Y.}~\bibnamefont{Zhang}},
  \bibinfo{author}{\bibfnamefont{T.-T.} \bibnamefont{Tang}},
  \bibinfo{author}{\bibfnamefont{C.}~\bibnamefont{Girit}},
  \bibinfo{author}{\bibfnamefont{Z.}~\bibnamefont{Hao}},
  \bibinfo{author}{\bibfnamefont{M.~C.} \bibnamefont{Martin}},
  \bibinfo{author}{\bibfnamefont{A.}~\bibnamefont{Zettl}},
  \bibinfo{author}{\bibfnamefont{M.~F.} \bibnamefont{Crommie}},
  \bibinfo{author}{\bibfnamefont{Y.~R.} \bibnamefont{Shen}}, \bibnamefont{and}
  \bibinfo{author}{\bibfnamefont{F.}~\bibnamefont{Wang}},
  \bibinfo{journal}{Nature} \textbf{\bibinfo{volume}{459}},
  \bibinfo{pages}{820} (\bibinfo{year}{2009}).

\bibitem[{\citenamefont{McCann}(2006)}]{mccann2006}
\bibinfo{author}{\bibfnamefont{E.}~\bibnamefont{McCann}},
  \bibinfo{journal}{Phys. Rev. B} \textbf{\bibinfo{volume}{74}},
  \bibinfo{pages}{161403} (\bibinfo{year}{2006}).

\bibitem[{\citenamefont{Nilsson et~al.}(2008)\citenamefont{Nilsson,
  Castro~Neto, Guinea, and Peres}}]{nilsson2008}
\bibinfo{author}{\bibfnamefont{J.}~\bibnamefont{Nilsson}},
  \bibinfo{author}{\bibfnamefont{A.~H.} \bibnamefont{Castro~Neto}},
  \bibinfo{author}{\bibfnamefont{F.}~\bibnamefont{Guinea}}, \bibnamefont{and}
  \bibinfo{author}{\bibfnamefont{N.~M.~R.} \bibnamefont{Peres}},
  \bibinfo{journal}{Phys. Rev. B} \textbf{\bibinfo{volume}{78}},
  \bibinfo{pages}{045405} (\bibinfo{year}{2008}).

\bibitem[{\citenamefont{Castro~Neto et~al.}(2009)\citenamefont{Castro~Neto,
  Guinea, Peres, Novoselov, and Geim}}]{castroneto2009}
\bibinfo{author}{\bibfnamefont{A.~H.} \bibnamefont{Castro~Neto}},
  \bibinfo{author}{\bibfnamefont{F.}~\bibnamefont{Guinea}},
  \bibinfo{author}{\bibfnamefont{N.~M.~R.} \bibnamefont{Peres}},
  \bibinfo{author}{\bibfnamefont{K.~S.} \bibnamefont{Novoselov}},
  \bibnamefont{and} \bibinfo{author}{\bibfnamefont{A.~K.} \bibnamefont{Geim}},
  \bibinfo{journal}{Rev. Mod. Phys.} \textbf{\bibinfo{volume}{81}},
  \bibinfo{pages}{109} (\bibinfo{year}{2009}).

\bibitem[{\citenamefont{Michetti et~al.}(2010)\citenamefont{Michetti, Recher,
  and Iannaccone}}]{michetti2010b}
\bibinfo{author}{\bibfnamefont{P.}~\bibnamefont{Michetti}},
  \bibinfo{author}{\bibfnamefont{P.}~\bibnamefont{Recher}}, \bibnamefont{and}
  \bibinfo{author}{\bibfnamefont{G.}~\bibnamefont{Iannaccone}},
  \bibinfo{journal}{Nano Lett.} \textbf{\bibinfo{volume}{10}},
  \bibinfo{pages}{4463} (\bibinfo{year}{2010}).

\bibitem[{\citenamefont{Barbier et~al.}(2009)\citenamefont{Barbier,
  Vasilopoulos, Peeters, and Pereira}}]{barbier2009}
\bibinfo{author}{\bibfnamefont{M.}~\bibnamefont{Barbier}},
  \bibinfo{author}{\bibfnamefont{P.}~\bibnamefont{Vasilopoulos}},
  \bibinfo{author}{\bibfnamefont{F.~M.} \bibnamefont{Peeters}},
  \bibnamefont{and} \bibinfo{author}{\bibfnamefont{J.~M.}
  \bibnamefont{Pereira}}, \bibinfo{journal}{Phys. Rev. B}
  \textbf{\bibinfo{volume}{79}}, \bibinfo{pages}{155402}
  (\bibinfo{year}{2009}).

\bibitem[{\citenamefont{Tedrow et~al.}(1986)\citenamefont{Tedrow, Tkaczyk, and
  Kumar}}]{tedrow1986}
\bibinfo{author}{\bibfnamefont{P.~M.} \bibnamefont{Tedrow}},
  \bibinfo{author}{\bibfnamefont{J.~E.} \bibnamefont{Tkaczyk}},
  \bibnamefont{and} \bibinfo{author}{\bibfnamefont{A.}~\bibnamefont{Kumar}},
  \bibinfo{journal}{Phys. Rev. Lett.} \textbf{\bibinfo{volume}{56}},
  \bibinfo{pages}{1746} (\bibinfo{year}{1986}).

\bibitem[{\citenamefont{Schmehl et~al.}(2007)\citenamefont{Schmehl,
  Vaithyanathan, Herrnberger, Thiel, Richter, Liberati, Heeg, Rockerath,
  Kourkoutis, Muhlbauer et~al.}}]{schmehl2007}
\bibinfo{author}{\bibfnamefont{A.}~\bibnamefont{Schmehl}},
  \bibinfo{author}{\bibfnamefont{V.}~\bibnamefont{Vaithyanathan}},
  \bibinfo{author}{\bibfnamefont{A.}~\bibnamefont{Herrnberger}},
  \bibinfo{author}{\bibfnamefont{S.}~\bibnamefont{Thiel}},
  \bibinfo{author}{\bibfnamefont{C.}~\bibnamefont{Richter}},
  \bibinfo{author}{\bibfnamefont{M.}~\bibnamefont{Liberati}},
  \bibinfo{author}{\bibfnamefont{T.}~\bibnamefont{Heeg}},
  \bibinfo{author}{\bibfnamefont{M.}~\bibnamefont{Rockerath}},
  \bibinfo{author}{\bibfnamefont{L.~F.} \bibnamefont{Kourkoutis}},
  \bibinfo{author}{\bibfnamefont{S.}~\bibnamefont{Muhlbauer}},
  \bibnamefont{et~al.}, \bibinfo{journal}{Nat. Mater.}
  \textbf{\bibinfo{volume}{6}}, \bibinfo{pages}{882} (\bibinfo{year}{2007}).

\bibitem[{\citenamefont{Swartz et~al.}(2010)\citenamefont{Swartz, Ciraldo,
  Wong, Li, Han, Lin, Mack, Shi, Awschalom, and Kawakami}}]{swartz2010}
\bibinfo{author}{\bibfnamefont{A.~G.} \bibnamefont{Swartz}},
  \bibinfo{author}{\bibfnamefont{J.}~\bibnamefont{Ciraldo}},
  \bibinfo{author}{\bibfnamefont{J.~J.~I.} \bibnamefont{Wong}},
  \bibinfo{author}{\bibfnamefont{Y.}~\bibnamefont{Li}},
  \bibinfo{author}{\bibfnamefont{W.}~\bibnamefont{Han}},
  \bibinfo{author}{\bibfnamefont{T.}~\bibnamefont{Lin}},
  \bibinfo{author}{\bibfnamefont{S.}~\bibnamefont{Mack}},
  \bibinfo{author}{\bibfnamefont{J.}~\bibnamefont{Shi}},
  \bibinfo{author}{\bibfnamefont{D.~D.} \bibnamefont{Awschalom}},
  \bibnamefont{and} \bibinfo{author}{\bibfnamefont{R.~K.}
  \bibnamefont{Kawakami}}, \bibinfo{journal}{Appl. Phys. Lett.}
  \textbf{\bibinfo{volume}{97}}, \bibinfo{pages}{112509}
  (\bibinfo{year}{2010}).

\bibitem[{\citenamefont{Steeneken et~al.}(2002)\citenamefont{Steeneken, Tjeng,
  Elfimov, Sawatzky, Ghiringhelli, Brookes, and Huang}}]{steeneken2002}
\bibinfo{author}{\bibfnamefont{P.~G.} \bibnamefont{Steeneken}},
  \bibinfo{author}{\bibfnamefont{L.~H.} \bibnamefont{Tjeng}},
  \bibinfo{author}{\bibfnamefont{I.}~\bibnamefont{Elfimov}},
  \bibinfo{author}{\bibfnamefont{G.~A.} \bibnamefont{Sawatzky}},
  \bibinfo{author}{\bibfnamefont{G.}~\bibnamefont{Ghiringhelli}},
  \bibinfo{author}{\bibfnamefont{N.~B.} \bibnamefont{Brookes}},
  \bibnamefont{and} \bibinfo{author}{\bibfnamefont{D.-J.} \bibnamefont{Huang}},
  \bibinfo{journal}{Phys. Rev. Lett.} \textbf{\bibinfo{volume}{88}},
  \bibinfo{pages}{047201} (\bibinfo{year}{2002}).

\bibitem[{\citenamefont{Santos et~al.}(2008)\citenamefont{Santos, Moodera,
  Raman, Negusse, Holroyd, Dvorak, Liberati, Idzerda, and
  Arenholz}}]{santos2008}
\bibinfo{author}{\bibfnamefont{T.~S.} \bibnamefont{Santos}},
  \bibinfo{author}{\bibfnamefont{J.~S.} \bibnamefont{Moodera}},
  \bibinfo{author}{\bibfnamefont{K.~V.} \bibnamefont{Raman}},
  \bibinfo{author}{\bibfnamefont{E.}~\bibnamefont{Negusse}},
  \bibinfo{author}{\bibfnamefont{J.}~\bibnamefont{Holroyd}},
  \bibinfo{author}{\bibfnamefont{J.}~\bibnamefont{Dvorak}},
  \bibinfo{author}{\bibfnamefont{M.}~\bibnamefont{Liberati}},
  \bibinfo{author}{\bibfnamefont{Y.~U.} \bibnamefont{Idzerda}},
  \bibnamefont{and} \bibinfo{author}{\bibfnamefont{E.}~\bibnamefont{Arenholz}},
  \bibinfo{journal}{Phys. Rev. Lett.} \textbf{\bibinfo{volume}{101}},
  \bibinfo{pages}{147201} (\bibinfo{year}{2008}).

\bibitem[{\citenamefont{M\"uller et~al.}(2009)\citenamefont{M\"uller, Miao, and
  Moodera}}]{muller2009}
\bibinfo{author}{\bibfnamefont{M.}~\bibnamefont{M\"uller}},
  \bibinfo{author}{\bibfnamefont{G.-X.} \bibnamefont{Miao}}, \bibnamefont{and}
  \bibinfo{author}{\bibfnamefont{J.~S.} \bibnamefont{Moodera}},
  \bibinfo{journal}{J. Appl. Phys.} \textbf{\bibinfo{volume}{105}},
  \bibinfo{pages}{07C917} (\bibinfo{year}{2009}).

\bibitem[{\citenamefont{Merkulov et~al.}(1999)\citenamefont{Merkulov, Yakovlev,
  Keller, Ossau, Geurts, Waag, Landwehr, Karczewski, Wojtowicz, and
  Kossut}}]{merkulov1999}
\bibinfo{author}{\bibfnamefont{I.~A.} \bibnamefont{Merkulov}},
  \bibinfo{author}{\bibfnamefont{D.~R.} \bibnamefont{Yakovlev}},
  \bibinfo{author}{\bibfnamefont{A.}~\bibnamefont{Keller}},
  \bibinfo{author}{\bibfnamefont{W.}~\bibnamefont{Ossau}},
  \bibinfo{author}{\bibfnamefont{J.}~\bibnamefont{Geurts}},
  \bibinfo{author}{\bibfnamefont{A.}~\bibnamefont{Waag}},
  \bibinfo{author}{\bibfnamefont{G.}~\bibnamefont{Landwehr}},
  \bibinfo{author}{\bibfnamefont{G.}~\bibnamefont{Karczewski}},
  \bibinfo{author}{\bibfnamefont{T.}~\bibnamefont{Wojtowicz}},
  \bibnamefont{and} \bibinfo{author}{\bibfnamefont{J.}~\bibnamefont{Kossut}},
  \bibinfo{journal}{Phys. Rev. Lett.} \textbf{\bibinfo{volume}{83}},
  \bibinfo{pages}{1431} (\bibinfo{year}{1999}).

\bibitem[{\citenamefont{Smirnov}(2001)}]{smirnov2001}
\bibinfo{author}{\bibfnamefont{B.~M.} \bibnamefont{Smirnov}},
  \bibinfo{journal}{Phys.-Usp.} \textbf{\bibinfo{volume}{44}},
  \bibinfo{pages}{221} (\bibinfo{year}{2001}).

\bibitem[{\citenamefont{Tokuyasu et~al.}(1988)\citenamefont{Tokuyasu, Sauls,
  and Rainer}}]{tokuyasu1988}
\bibinfo{author}{\bibfnamefont{T.}~\bibnamefont{Tokuyasu}},
  \bibinfo{author}{\bibfnamefont{J.~A.} \bibnamefont{Sauls}}, \bibnamefont{and}
  \bibinfo{author}{\bibfnamefont{D.}~\bibnamefont{Rainer}},
  \bibinfo{journal}{Phys. Rev. B} \textbf{\bibinfo{volume}{38}},
  \bibinfo{pages}{8823} (\bibinfo{year}{1988}).

\end{thebibliography}
\end{document}